\documentclass[aps,prd,twocolumn,
preprintnumbers,amsmath,amssymb,
epsf,superscriptaddress,groupedaddress,tightenlines,nofootinbib]{revtex4}
\usepackage{epsfig,graphicx}

\usepackage{graphicx}
\usepackage{amsmath}
\usepackage{bm}

\newcommand{\beqa}{\begin{eqnarray}}
\newcommand{\eeqa}{\end{eqnarray}}
\newcommand{\beq}{\begin{equation}}
\newcommand{\eeq}{\end{equation}}
\newcommand{\bsp}{\begin{split}}
\newcommand{\esp}{\end{split}}
\newcommand{\bal}{\begin{align}}
\newcommand{\eal}{\end{align}}

\begin{document}


\vskip 0.5cm
\vspace*{10pt}
\title{Constraint on $\bar\rho, \bar\eta$ from $B\to K^* \pi$}
\def\addbnl{Physics Department, Brookhaven National Laboratory, Upton,
New York 11973}
\def\addtech{On sabbatical leave from the Physics Department, 
Technion--Israel Institute of Technology,
Haifa 32000, Israel}
\def\addslac{Stanford Linear Accelerator Center, Stanford University,
Stanford, CA 94309}
\def\addCERN{Theory Division, Department of Physics, CERN, CH-1211
Geneva 23,
Switzerland}
\def\addFMF{Faculty of mathematics and physics, University of Ljubljana,
Jadranska 19, 1000
Ljubljana, Slovenia}
\def\addIJS{J. Stefan Institute, Jamova 39, P.O. Box 3000, 1001
Ljubljana, Slovenia}
\def\addBucarest{National Institute for Physics and Nuclear Engineering,
Department of Particle Physics, 077125 Bucharest, Romania}

\author{Michael Gronau}\thanks{\addtech}\affiliation{\addslac}
\author{Dan Pirjol}\affiliation{\addBucarest}
\author{Amarjit Soni}\affiliation{\addbnl}
\author{Jure Zupan}
\affiliation{\addCERN}\affiliation{\addFMF}\affiliation{\addIJS}

\begin{abstract} \vspace*{18pt}
A linear relation between Cabibbo-Kobayashi-Maskawa quark mixing
parameters, $\bar\eta= \tan\Phi_{3/2}(\bar\rho-0.24\pm 0.03)$, 
involving a $1\sigma$ range for $\Phi_{3/2}$, $20^\circ < \Phi_{3/2} < 115^\circ$,  
is obtained from $B^0\to K^*\pi$ amplitudes measured in Dalitz plot
analyses of $B^0\to K^+\pi^-\pi^0$ and $B^0(t)\to K_S\pi^+\pi^-$. This relation
is consistent within the large error on $\Phi_{3/2}$ with other CKM constraints.
We discuss the high sensitivity of this method to a new physics contribution 
in the $\Delta S=\Delta I=1$ amplitude.
\end{abstract}
\maketitle

\section{Introduction}

Two anomalous features measured in  $b\to s$ penguin-dominated processes
have attracted substantial interest in recent years~\cite{Gronau:2007xg}:
(i) CP asymmetries $\Delta S$ in 
$B^0\to K_S X$ decays~($X=\pi^0, \phi, \eta', \rho^0, \omega, K_SK_S, \pi^0K_S$) 
show a hint of systematic deviations from standard model predictions, and
(ii)  the pattern of direct CP asymmetries in $B\to K\pi$ decays is hard to explain
using dynamical approaches  based on $1/m_b$ expansion.
Are these merely statistical fluctuations, a sign of our inabilities to reliably calculate the relevant
observables, or are they first hints of new flavor-dependent CP-violating contributions
from new physics at a TeV scale?

In order to answer this question it is important to obtain precise model-independent 
constraints on the  CKM parameters  $\bar\rho$ and $\bar\eta$~\cite{Wolfenstein:1983yz}
using  penguin dominated $\Delta S=1$ $B$ decays. Comparing these constraints with CKM 
constraints which are not affected by New Physics (NP) in $\Delta S=1$ decays, e.g.,
the determination of $\gamma$ from  tree-dominated processes $B\to D^{(*)}K^{(*)}$~\cite{Gronau:1990ra}, may provide a test for the presence of NP in $b\to s$ penguin transitions.

In the present note we study a linear constraint in the ($\bar\rho, \bar\eta$) plane following from a combination of $B^0\to K^*\pi$ amplitudes.
The method proposed in~\cite{Ciuchini:2006kv}
and developed further in~\cite{Gronau:2006qn} will be summarized in Section II.
The necessary observables required for applying the method have been measured recently 
in Dalitz plot analyses of $B^0\to K^+\pi^-\pi^0$~\cite{Aubert:2007bs} and 
$B^0\to K_S\pi^+\pi^-$~\cite{:2007vi}. They will be used in Section III 
to determine the slope of the linear constraint, comparing this constraint with other 
CKM constraints. Section IV discusses the sensitivity of this test to New Physics effects, 
while Section V concludes.

\section{The method}

The main idea of the method~\cite{Ciuchini:2006kv,Gronau:2006qn}
 is studying $\Delta I=1$ combinations of  $B\to K^*\pi$
amplitudes which do not receive dominant contributions from QCD penguin operators,
and thus carry a weak phase $\gamma$ in the absence of electroweak penguin (EWP) terms. 
In the present note we focus our attention on the $I=3/2$ final state,
\beq\label{A3/2}
3A_{3/2} =  A(B^0\to K^{*+}\pi^-) + \sqrt{2}A(B^0\to K^{*0}\pi^0)~.
\eeq
In the absence of EWP  terms $\gamma$ would be given by
\beq\label{Phi3/2}
\gamma= \Phi_{3/2}\equiv -\frac{1}{2}\mbox{arg}\left(R_{3/2}\right )~,~~~~~~
R_{3/2}\equiv \frac{\bar A_{3/2}}{A_{3/2}}~,
\eeq
where $\bar A_{3/2}$ is the amplitude for charge-conjugated states.

The phase $\Phi_{3/2}$ can be obtained by measuring magnitudes and relative phases 
of $B^0\to K^{*+}\pi^-$ and $B^0\to K^{*0}\pi^0$ amplitudes and their charge-conjugates. 
The advantage of $B\to K^*\pi$ over $B\to K\pi$ decays is that $K^*\pi$ quasi--two-body states
occur in Dalitz plots of $B\to K\pi\pi$, where overlapping resonances permit determining 
both the magnitudes and relative phases of $B\to K^*\pi$ amplitudes.
In contrast, the relative phases of $B\to K\pi$ amplitudes cannot be measured directly.

The inclusion of EWP contributions modifies the expression for 
$R_{3/2}$ which becomes~\cite{Gronau:2006qn}
\beqa\label{R32}
R_{3/2} & = & e^{-2i[\gamma + {\rm arg}(1+\kappa)]}\frac{1+c^*_{\kappa}r_{3/2}}
{1+c_{\kappa}r_{3/2}}~,\\
\label{kappa}
\kappa  & \equiv &  -\frac{3}{2}\frac{C_9+C_{10}}{C_1+C_2}
\frac{V^*_{tb}V_{ts}}{V^*_{ub}V_{us}}~,~~c_{\kappa}\equiv \frac{1-\kappa}{1+\kappa}~,\\
r_{3/2} & \equiv & \frac{(C_1-C_2)\langle (K^*\pi)_{I=3/2}|{\cal O}_1-{\cal O}_2|B^0\rangle}
{(C_1+C_2)\langle (K^*\pi)_{I=3/2}|{\cal O}_1+{\cal O}_2|B^0\rangle}~.
\label{r3/2}
\eeqa
Here ${\cal O}_1\equiv (\bar bs)_{\rm V-A}(\bar uu)_{\rm V-A}$ and 
${\cal O}_2\equiv (\bar bu)_{\rm V-A}(\bar us)_{\rm V-A}$ are the V-A current-current 
operators.

The straight line $\bar\eta = \bar\rho\,\tan\Phi_{3/2}$, in the 
absence of EWP terms, is shifted by these contributions
along the $\bar\rho$ axis by a calculable finite
amount. The actual constraint becomes~\cite{Gronau:2006qn}
\beq\label{constraint}
\bar\eta = \tan\Phi_{3/2}\left [\bar\rho + C[1- 2{\rm Re}(r_{3/2})] + {\cal O}(r^2_{3/2})\right]~,
\eeq
where ($\lambda=0.227$)
\beq
C  \equiv \frac{3}{2}\frac{C_9+C_{10}}{C_1+C_2}\frac{1-\lambda^2/2}
{\lambda^2} = -0.27~.
\eeq

A finite positive shift of the straight line (\ref{constraint}) along the $\bar\rho$ axis, given by 
$-C=0.27$, is obtained using next to leading order values of Wilson coefficients $C_i$ at 
$\mu=m_b$~\cite{Buchalla:1995vs}. 
The theoretical error in this parameter is smaller than $1\%$. The complex parameter $r_{3/2}$ was
calculated in factorization, which gives a real result of the order of several
percent, $r_{3/2} \leq 0.05$~\cite{Ciuchini:2006kv}.

A similar but more conservative result is obtained  for $r_{3/2}$ by applying flavor SU(3) to corresponding $\Delta S=0$ decay amplitudes. Noting that the operators in the numerator and denominator in (\ref{r3/2}) transform as $\bf 6$ and $\bf\overline{15}$ of SU(3), one 
finds~\cite{Gronau:2006qn},
\beq
\begin{split}
r_{3/2}&=\frac{|\sqrt{{\cal B}(\rho^+\pi^0)} - \sqrt{{\cal B}(\rho^0\pi^+)}|}
{\sqrt{{\cal B}(\rho^+\pi^0)} + \sqrt{{\cal B}(\rho^0\pi^+)}} \\
&= 0.054 \pm 0.045 \pm 0.023~.
\label{r_{3/2}}
\end{split}
\eeq
The first error is experimental. The second error is due to SU(3) breaking,  small
$\Delta S=0$ penguin amplitudes and small strong phase difference between $B\to \rho\pi$
decay amplitudes which are neglected. 

We have assumed that SU(3) breaking in ratios of $\Delta S=1$ amplitudes and 
corresponding $\Delta S=0$ amplitudes introduces an uncertainty of 30$\%$ in these 
ratios. The $B\to \rho\pi$ phase difference is expected to be suppressed
by $1/m_b$ and $\alpha_s(m_b)$~\cite{Beneke:1999br,Bauer:2004tj}. Indeed,
evidence for a small phase difference is provided by an isospin pentagon relation obeyed by
measured $B\to \rho\pi$ amplitudes~\cite{Gronau:2006qn}.
The error in (\ref{constraint}) from neglecting this small strong phase difference is negligible 
because ${\rm Re}(r_{3/2})$ depends quadratically on this phase.
We will use the calculation (\ref{r_{3/2}}) for $r_{3/2}$ which is more conservative than 
the one using factorization. Combining in quadrature the two errors in $r_{3/2}$, the constraint (\ref{constraint}) becomes
\beq\label{constraint-num}
\bar\eta = \tan\Phi_{3/2}\left[\bar\rho -0.24\pm0.03\right]~.
\eeq
The dominant uncertainty in this linear constraint originates in $r_{3/2}$.

\begin{figure}
\includegraphics[height=6cm, width=6cm]{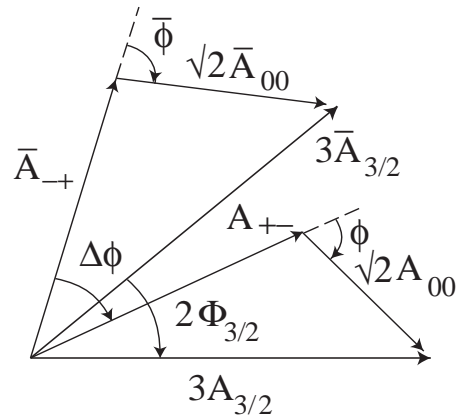}
\caption{Geometry for Eq.~(\ref{A3/2}) and its charge-conjugate,
using notations $A_{+-} \equiv A(B^0 \to K^{*+}\pi^-),
A_{00} =  A(B^0 \to K^{*0}\pi^0)$ and similar notations for charge-conjugated
modes.}
\label{triangle}
\end{figure}

Eq.~(\ref{R32}) and a real value of $r_{3/2}$ imply $|R_{3/2}|=1$.
The strong phase of $r_{3/2}$ is expected to be suppressed by $1/m_b$ and 
$\alpha_s(m_b)$~\cite{Beneke:1999br,Bauer:2004tj}. Using (\ref{r_{3/2}}) we take
 \beq\label{r32}
 |r_{3/2}|<0.11,~~~~~|{\rm arg}(r_{3/2})| < 30^\circ~,
 \eeq
leading to the bounds
\beq\label{constrainedR}
0.8 < |R_{3/2}| < 1.2~.
\eeq

\section{Determining $\Phi_{3/2}$}

The phase $\Phi_{3/2}$ can be determined by
measuring the magnitudes and relative phases of the  
$B^0\to K^{*+}\pi^-,~B^0\to K^{*0}\pi^0$ amplitudes and their charge-conjugates.
A graphical representation of the triangle relation Eq.~(\ref{A3/2}) and its
charge conjugate is given in Fig.~\ref{triangle}.

The above four magnitudes of amplitudes and the two relative phases, 
$\phi\equiv {\rm arg}[A(B^0\to K^{*0}\pi^0)/A(B^0\to K^{*+}\pi^-)]$
and $\bar\phi\equiv {\rm arg}[A(\bar B^0\to \bar K^{*0}\pi^0)/A(\bar B^0\to K^{*-}\pi^+)]$,
determine the two triangles separately. These quantities have been measured recently
in a Dalitz plot analysis of $B^0\to K^+\pi^-\pi^0$ and its charge-conjugate~\cite{Aubert:2007bs}.
The relative phase $\Delta\phi\equiv {\rm arg}[A(B^0\to K^{*+}\pi^-)/A(\bar B^0\to K^{*-}\pi^+)]$,
which  fixes the relative orientation of the two triangles, has been measured in a
 time-dependent Dalitz plot analysis of $B^0\to K_S\pi^+\pi^-$~\cite{:2007vi}.
 
\begin{table}[t]
\begin{ruledtabular}
\begin{tabular}{lll}
Mode & Branching ratio & $A_{CP}$\\
\hline
$K^{*+}\pi^-$ &$10.4\pm 0.9$ & $-0.14\pm 0.12$\\
$K^{*0}\pi^0$ &$3.6\pm 0.9$ & $-0.09 \pm 0.24$\\
\end{tabular}
\end{ruledtabular}
\caption{Branching ratios in units of $10^{-6}$ and CP asymmetries
in $B^0\to K^*\pi$~\cite{Aubert:2007bs,Barberio:2007cr}.}
\label{tab:1}
\end{table}

Table I  quotes CP-averaged branching ratios and CP
asymmetries for $B^0\to K^{*+}\pi^-,~B^0\to K^{*0}\pi^0$ using 
Refs.~\cite{Aubert:2007bs} and \cite{Barberio:2007cr}. 
A value $\Delta\phi=(-164\pm 30.7)^\circ$ was measured in 
$B^0(t)\to K_S\pi^+\pi^-$~\cite{:2007vi}. The experimental situation is less clear
for the phases $\phi$ and $\bar\phi$,  measured recently in an amplitude analysis 
performed for $B^0\to K^+\pi^-\pi^0$ and its charge-conjugate~\cite{Aubert:2007bs}.

\begin{figure}[h]
\includegraphics[width=8cm, height=5cm]{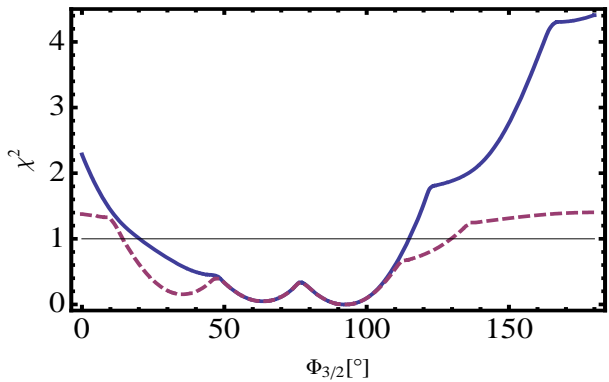}
\caption{$\chi^2$ dependence on $\Phi_{3/2}$ for unconstrained $|R_{3/2}|$ (broken 
purple line) and for $0.8 < |R_{3/2}| <1.2$ (solid blue line). A black horizontal line at 
$\chi^2=1$ defines $1\sigma$ ranges for $\Phi_{3/2}$.}
\label{chi2Phi32}
\end{figure}

In order to calculate the $\chi^2$ dependence on $\Phi_{3/2}$
we use the $\chi^2$ dependence on $\phi$ and $\bar \phi$ given in 
Ref.~\cite{Aubert:2007bs}, assuming gaussian errors 
for $\Delta\phi$ and for branching ratios and CP asymmetries in 
$B^0\to K^{*+}\pi^-$ and $B^0\to K^{*0}\pi^0$. Potential correlations between $\phi, \bar\phi$ 
and branching ratios and asymmetries are neglected. Two resulting $\chi^2$ plots as function
of $\Phi_{3/2}$ are shown in Fig.~\ref{chi2Phi32}. The broken purple curve corresponds to an
unconstrained $|R_{3/2}|$, while the solid blue curve is obtained by imposing the bounds
(\ref{constrainedR}), expected to hold in the Standard Model.
The latter curve defines a $1\sigma$ range,
\beq\label{rangePhi32}
20^\circ < \Phi_{3/2} < 115^\circ~.
\eeq 

\begin{figure}
\includegraphics[width=8cm, height=5cm]{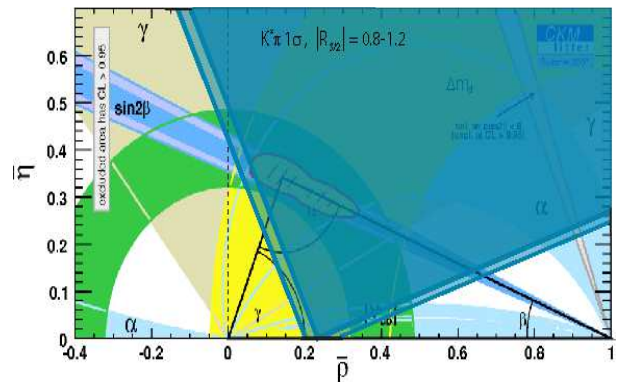}
\caption{Constraint in the $\bar \rho-\bar \eta$ plane 
following from Eqs.~(\ref{constraint-num}) and (\ref{rangePhi32}). 
The dark  shaded region marked $K^*\pi\,1\sigma$ corresponds to the experimental 
error on $\Phi_{3/2}$ given by the $1\sigma$ range (\ref{rangePhi32}), 
while the light shaded region includes also the error on $r_{3/2}$ \eqref{r_{3/2}}.
Also shown are CKMfitter constraints obtained using $|V_{ub}|/|V_{cb}|, \beta, \alpha,\gamma$ 
and $\Delta m_d$~\cite{CKMfitter}.}
\label{ckm}
\end{figure}

Fig.~\ref{ckm} shows the linear constraint (\ref{constraint-num}) with the large range 
of slopes (\ref{rangePhi32}) overlaid on CKMFitter results 
following from~\cite{Barberio:2007cr,CKMfitter} $|V_{ub}|/|V_{cb}| = 0.086 \pm 0.009$, 
obtained in semileptonic $B$ decays, and values  $\beta=(21.5\pm 1.0)^\circ$,  $\alpha=
(88\pm 6)^\circ$ and $\gamma = (53^{+15}_{-18}\pm 3 \pm 9)^{\circ}$~\cite{Poluektov:2006ia}, 
obtained in $B\to J/\psi K_S$, $B\to \pi\pi, \rho\rho, \rho\pi$ and $B^+\to D^{(*)}K^{(*)+}$, 
respectively. 
The small theoretical error in the $B\to K^*\pi$ constraint  [$\pm 0.03$ in Eq.
(\ref{constraint-num})] is described by the difference between dark and light 
shaded regions in Fig. \ref{ckm}. 
The large experimental error in $\Phi_{3/2}$ originates to a large extent in 
ambiguities in $\phi$ and $\bar\phi$ measured in $B^0\to K^+\pi^-\pi^0$, 
using an integrated luminosity on the $\Upsilon(4S)$ of only about 
200 fb$^{-1}$~\cite{Aubert:2007bs}. This error is expected to be reduced 
considerably by analyses based on higher up-to-date and future luminosities.  

\section{Sensitivity to New Physics}
As has already been stressed, new physics (NP) $\Delta S=1$ contributions may lead to an 
inconsistency between the linear constraint (\ref{constraint}) in
penguin dominated $B\to K^*\pi$ decays and values of $|V_{ub}|/|V_{cb}|, \beta, 
\alpha$ and $\gamma$ obtained in the above-mentioned processes.
The constraint (\ref{constraint}) is affected by $\Delta I=1$ NP operators, while 
NP contributions from potential $\Delta I=0$ operators drop out. A general discussion of ways for 
distinguishing between NP in $\Delta I=0$ and $\Delta I=1$ $b\to s$ transitions can be 
found in Ref.~\cite{Gronau:2007ut}.

The $I=3/2$ amplitude consists of complex tree and EWP terms, $T$ and $P_{EW}$,  
both of which involve strong phases,
\beq\label{structure}
A_{3/2} = Te^{i\gamma} - P_{EW}~.
\eeq
The ratio~\cite{Gronau:2006qn}
\beq\label{ratioSM}
\frac{P_{EW}}{T}  =  |\kappa|\frac{1-r_{3/2}}{1+r_{3/2}}
\eeq
involves the parameter $\kappa$ defined in (\ref{kappa}), 
which has some dependence on CKM matrix elements whose central values correspond 
to $|\kappa|\simeq0.66$. 

Allowing for a NP term $A_{NP}\exp(i\psi)$, where $A_{NP}$ involves a CP conserving 
strong phase while $\psi$ is a new CP-violating phase,  the $\Delta I=1$ amplitude becomes
\beq\label{NPdef1}
A_{3/2} = Te^{i\gamma} - P_{EW} + A_{NP} e^{i\psi}~.
\eeq
The NP term can  be reabsorbed quite generally in redefined tree and 
electroweak penguin-like contributions, $\bar T$ and $\bar P_{EW}$,  without
changing the structure (\ref{structure})~\cite{Botella:2005ks},
\beq\label{NPstructure1}
A_{3/2} = \bar Te^{i\gamma} - \bar P_{EW}~.
\eeq
Here
\begin{eqnarray}\label{RPI1}
&& \bar T = T + A_{NP} \frac{\sin \psi}{\sin \gamma}~, \nonumber\\
&& \bar P_{EW} = P_{EW} + A_{NP}
\frac{\sin (\psi-\gamma)}{\sin \gamma}~.
\end{eqnarray}

The amplitudes $\bar T$ and $\bar P_{EW}$ can be used to define a  complex 
parameter $\bar r$ in analogy to Eq.~(\ref{ratioSM}),
\beq
\frac{\bar P_{EW}}{\bar T} = |\kappa|\frac{1-\bar r}{1+\bar r}~.
\eeq
Thus, the parameter $\bar r$ replaces $r_{3/2}$ in the expression 
(\ref{R32}) for $R_{3/2}$. Values of $\bar r$ outside the range (\ref{r32}) lead for 
most such values (unless ${\rm arg}(\bar r)$ is small) to a violation of the bounds 
(\ref{constrainedR}). {\em This would be likely evidence for New Physics.}

\begin{figure}[t]
\includegraphics[width=8cm, height=5cm]{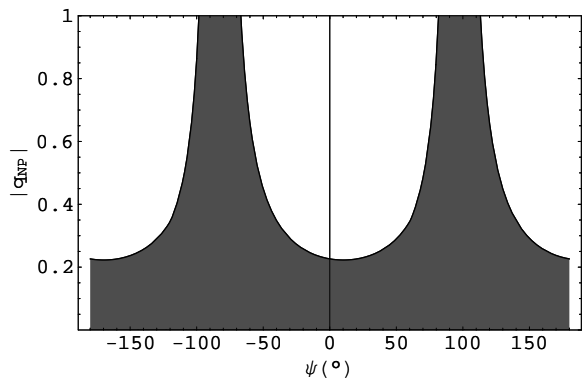}
\caption{Values of $|q_{NP}|$ and $\psi$ providing a signal for NP 
(at $\gamma=60^\circ$) are 
given by points outside the dark area, which is obtained by 
requiring values of $r_{3/2}$ and $\bar r$ in the range (\ref{r32}).}
 \label{qNP_vs_psi2}
\end{figure}
  
A criterion for the sensitivity of the method to observing a NP amplitude 
is provided by requiring that $\bar r$ lies outside the range 
 of values (\ref{r32}) allowed for $r_{3/2}$. Because of these small values 
 this criterion is expected to hold also for values of $A_{NP}$ which are small
 relative to $T$ and $P_{EW}$. 
 An exception is a singular case 
 where the weak phases $\psi$ and $\gamma$ are related by 
 \beq\label{singular}
 \frac{\sin(\psi-\gamma)}{\sin\psi} = \frac{P_{EW}}{T}~,
 \eeq
 for which $\bar P_{EW}/\bar T=P_{EW}/T$ is independent of $A_{NP}$. 
In the following discussion we will assume a value $\gamma=60^\circ$.
 
Denoting $q_{NP}=A_{NP}/P_{EW}$, 
we plot in the dark area in Fig.~\ref{qNP_vs_psi2} points corresponding to values of 
$|q_{NP}|$ and $\psi$, for which both $r_{3/2}$ and $\bar r$ are in the range (\ref{r32}).
The region outside this area, including for most values of $\psi$  rather small values of 
$|q_{NP}|$, $|q_{NP}|\sim 0.3$, implies a high sensitivity to an observable NP amplitude. 
 The spikes around $\psi \sim \pm 90^{\circ}$, implying very low sensitivity, correspond to 
 solutions of (\ref{singular}) and nearby lying values of $\psi$. 
 
\section{Conclusion}
Magnitudes and phases of $B^0\to K^*\pi$ decay amplitudes, extracted in Dalitz plot analyses
for $B^0\to K^+\pi^-\pi^0$ and $B^0\to K_S\pi^+\pi^-$, are used for obtaining the linear constraint
(\ref{constraint-num}) in the $\bar\rho,\bar\eta$ plane, 
where $\Phi_{3/2}$ lies in a $1\sigma$ range (\ref{rangePhi32}). 
This constraint is consistent with other
CKM constraints which are unaffected by NP $\Delta S=1$ operators. The dominant error 
in the slope of the straight line  is 
purely experimental, while a much smaller theoretical uncertainty occurs in a parallel shift
along the $\bar\rho$ axis. This small theoretical uncertainty is shown to imply in principle 
a high sensitivity to a New Physics $\Delta S=1, \Delta I=1$ amplitude.

\medskip
We thank Jacques Chauveau, Mathew Graham, Sebastian Jaeger, 
Jose Ocariz  and Soeren Prell for useful discussions. 
We are indebted to  Jacques Chauveau and Jose Ocariz 
for providing us with numerical $\chi^2$ dependence on $\phi$ and $\bar\phi$, and to 
Stephane T' Jampens for providing CKM constraints 
for Fig.~\ref{ckm}. The work of M. G. and A. S.
was supported in part by the US Department of Energy under contracts DE-AC02-76SF00515,
and DE-AC02-98CH10886, respectively. The work of J. Z. is supported in part by the European Commission RTN network, Contract No.~MRTN-CT-2006-035482 (FLAVIAnet) and by the 
Slovenian Research Agency.

\medskip
{\bf Note added}: After publication of this paper in Phys. Rev. D {\bf 77},
057504 (2008) the results of Ref. [6] were corrected. We updated our
analysis in a separate addendum.


\end{document}